\documentclass[final,5p,times,twocolumn]{elsarticle}
\usepackage{algorithm}
\usepackage{algpseudocode}
\usepackage{multirow}

\usepackage{amssymb}
\usepackage{amsthm}

\journal{The Journal of Systems and Software}

\begin{document}

\begin{frontmatter}



\title{A Quantitative Model for Predicting \\ Cross-application Interference in Virtual Environments}

\author[ic]{Maicon Melo Alves\corref{cor1}}
\ead{mmelo@ic.uff.br}
\author[ic]{L{\'u}cia Maria de Assump\c{c}\~ao Drummond}
\ead{lucia@ic.uff.br}

\cortext[cor1]{Corresponding author}

\address[ic]{Instituto de Computa\c{c}\~ao, Universidade Federal Fluminense, Niter{\'o}i, Brazil}

\begin{abstract}

Cross-application interference can affect drastically performance of HPC applications when running in clouds. This problem is caused by concurrent access performed by co-located applications to shared and non-sliceable resources such as cache and memory. In order to address this issue, some works adopted a qualitative approach that does not take into account the amount of access to shared resources. In addition, a few works, even considering the amount of access, evaluated just the SLLC access contention as the root of this problem. However, our experiments revealed that interference is intrinsically related to the amount of simultaneous access to shared resources, besides showing that another shared resources, apart from SLLC, can also influence the interference suffered by co-located applications. In this paper, we present a quantitative model for predicting cross-application interference in virtual environments. Our proposed model takes into account the amount of simultaneous access to SLLC, DRAM and virtual network, and the similarity of application's access burden to predict the level of interference suffered by applications when co-located in a same physical machine. Experiments considering a real petroleum reservoir simulator and applications from HPCC benchmark showed that our model reached an average and maximum prediction errors around 4\% and 12\%, besides achieving an error less than 10\% in approximately 96\% of all tested cases.

\end{abstract}


\end{frontmatter}

\section{Introduction}
\label{sec_introduction}

Cloud computing is emerging as a promising alternative to execute HPC (High Performance Computing) applications. 
 This new computational paradigm provides some attractive advantages when compared with a dedicated infrastructure, such as rapid provisioning of resources and significant reduction in operating costs related to energy, software licence and hardware obsolescence \cite{gholami2016cloud} \cite{dehury2016149} \cite{alves2014} \cite{elgazzar201664} \cite{kumar2016virtual} \cite{tsuruoka2016cloud} \cite{atif2016breaking}. 
 
However, some challenges must be overcome to bridge the gap between performance provided by a dedicated infrastructure and the one supplied by clouds. Overheads introduced by virtualization layer, hardware heterogeneity and   low latency networks, for example,  affect negatively  the performance of HPC applications  when executed  in clouds  \cite{alves2014} \cite{gupta2014} \cite{nanos2014xen2mx} \cite{chen2015profiling} \cite{lin20122593}. 
In addition, cloud providers usually adopt  resource sharing policies that can worse even more  HPC applications performances. Typically,  one physical server can host many virtual machines holding distinct applications \cite{shirvani2016server}, that may contend for shared and non-sliceable resources like cache and main memory \cite{gholami2016cloud} \cite{gupta2013hp} \cite{gupta2013hpc} \cite{jin2015ccap} \cite{mury2014concurrency} \cite{rameshan2016role},  reducing significantly their performances.

In order to address this  problem,  in \cite{mury2014concurrency} a classification based on the \textit{Thirteen Dwarfs}, which categorizes an application from its computational method,  was used to decide which  Dwarf classes could be co-located in a same physical machine without interference, i.e.,  keeping their original performances.   Another work,  \cite{jin2015ccap}, classifies HPC applications according to  their cache access pattern. So,  applications are classified within three SLLC access classes called: (i) cache-pollution, (ii) cache-sensitive and (iii) cache-friendly. From such classification, they claimed that a cache-pollution application, which presents weak-locality and large cache working set, should be primarily co-located with a cache-friendly application in order to alleviate cross-interference.

Those approaches, called here qualitative, because consider general characteristics of HPC applications, do not   determine precisely the cross-application interference. Concerning the  Dwarfs classification proposal, our experiments showed that the applications PTRANS and DGEMM\footnote{PTRANS and DGEMM belong to the set of applications provided by the High Performance Computing Challenge (HPCC) benchmark}, belonging to the same Dwarf class,  Dense Linear Algebra, presented distinct interference levels when they were co-located with themselves. Furthermore, our results also revealed that PTRANS presented distinct interference levels in face of instances of different sizes, allowing to assert that the  interference level can vary with the amount of data computed by the  application.  Also, in \cite{jin2015ccap}, two cache-friendly applications,  EP and IS\footnote{EP and IS belong to the set of applications provided by the Nas Parallel Benchmark (NPB)}, presented distinct interference levels when co-located with the same cache-pollution application, CG. Although both applications present a compatible cache access behavior, IS performed a higher number of memory references per second \cite{gupta2013hpc} what could explain why IS presented a higher interference level than EP.

Those results indicate  that an approach that considers the  amount of accesses to shared resources, could be a better strategy to determine more precisely the interference among applications co-located in a same physical machine. 
 In this context, in \cite{gupta2013hpc}  the relation between SLLC access burden and resulting cross-application interference was investigated.  Although that work is a step forward when compared with the others, our experiments showed that other shared resources, besides SLLC, can also influence the  interference level of  co-located applications and should be considered as well. 
  Moreover, all those related works  only indicate   whether applications should be or not co-located, but  does not quantify the interference level between them.     


In addition, our experimental tests revealed that the interference can also be influenced by the similarity of application's access burden. When applications co-located in the same host present a high level of access burden similarity for a given shared resource, they evenly compete for this resource which, in turn, leverages the level of interference suffered by these applications. Thus, besides the amount of simultaneous access performed to shared resources, the similarity of application's access burden should also be considered when investigating the cause of interference.   

In this paper, we propose a quantitative model for predicting interference among applications co-located in a same physical machine by considering the amount of  simultaneous accesses to  shared  resources and the similarity of application's access burden.  In order to predict that interference level, our proposed model considers the effect that SLLC, DRAM and virtual network  concurrent accesses impose in cross-application interference. Those three resources  are considered particularly critical because (i) SLLC and DRAM are shared among  cores of a processor and, (ii) virtual network, although not a hardware resource, is emulated by the hypervisor which is a central component shared by all virtual machines \cite{xu2010cache} \cite{albericio2013reuse}.

In order to build this prediction model,  at first we proposed an application template from which synthetic applications with distinct access patterns to each shared resource are generated. We measured the interferences, when those applications were executed concurrently, and  generated  an interference dataset containing these results.  Then, our quantitative model was built by applying the Multiple Regression Analysis (MRA), one of the most widely statistical procedures used for treating multivariate problems,  over that interference dataset. Note that MRA is a powerful and well-known  technique to build a model capable of  predicting an  unknown value of a response variable from the known values of multiple independent variables \cite{sudevalayam2013} \cite{atici2011} \cite{hair2006multivariate} \cite{mason1991collinearity} \cite{ngo2012steps}.

We validated the prediction model by using a real petroleum reservoir simulator, called Multiphase Filtration Transport Simulator (MUFITS)  \cite{afanasyev2016validation} \cite{coco2016numerical}, and applications from a well-known computing benchmark, the High Performance Computing Challenge (HPCC) \cite{dongarra2004introduction} \cite{lin20122593} \cite{jackson2010performance}. Those applications were executed with different instances  as input  and results showed that our model predicted cross-application interference with a maximum and average errors of 12,03\% and 4,06\%, respectively. Besides that, the prediction error was less than 10\% in 95,56\%  for all tested cases.  

Thus, the main contributions of this work are the following:

\begin{itemize}
    \item An interference prediction quantitative model that takes into account the amount of simultaneous access to SLLC, DRAM and virtual network, and the similarity of application's access burden to predict the interference level suffered by co-located applications. \item An application template able to generate applications with distinct access rates to SLLC, DRAM and virtual network. 
    \item A systematic evaluation of interference suffered by co-located applications before distinct levels of access to shared resources.  
    \item An evaluation of model precision by considering a real reservoir petroleum simulator and applications from a well-known HPC benchmark. 
    
\end{itemize}

At last, it is worth mentioning that the recent growth in the number of cores available in newer processors can increase even more the number of applications hosted in a same physical machine\footnote{In March of 2016, Intel\textsuperscript{\textregistered} launched processor E5-2600 v4, codename Broadwell-EP, which is endowed with 22 cores. Thus, a physical server that supports two of these processors, such as Supermicro\textsuperscript{\textregistered} 1028U-TN10RT+, provides 44 cores in total. }. 
In order to run  HPC applications in those systems, Virtual Machine Placement (VMP) strategies  will have to include efficient solutions for the cross-application interference problem.

The remainder of this paper is organized as follows. Section 2 presents works from the related literature. Section 3 presents the experiments executed to generate an interference dataset. Section 4 describes the model for predicting interference.  The Model validation is discussed in Section 5. Finally, conclusions and future work are presented in Section 6.

\section{Related Works}
\label{sec:related}

In this section, related works that investigated or just introduced the cross-application interference problem are presented. 

Some works presented the cross-application interference, but did not propose a solution to determine or alleviate this problem. Yokoyama \cite{yokoyama2015} and Basto \cite{basto2015} accomplished interference experiments by using benchmark applications in order to  generate a static matrix of interferences. Such matrix was further used as the basis for their proposed interference-aware VMP strategies. They evaluated their VMP strategies just using applications previously used in interference experiments. Thus, unlike our work, those papers did not present a solution to determine interference among co-located applications. 

Other works, besides introducing interference, proposed a naive or limited approach to explain this problem. Jersak \textit{et al.} \cite{jersak2016performance} devised a simple interference model to be used as a proof of concept with its proposed VMP strategy. In such model, the level of interference is defined as a function of the  number of virtual machines co-executing in physical machine. So, the model considers that the higher the number of virtual machines, the higher interference level will be. This naive strategy is not able to determine interference since our experimental results showed that, for the same number of virtual machines, the interference can vary drastically. This happens because interference is actually related to the amount of access to shared resources and not to the number of virtual machines being co-executed in host. 

Rameshan \textit{et al.} \cite{rameshan2014stay} proposed a mechanism to prevent latency sensitive applications from being adversely affected by best-effort batch applications when co-located in a same physical machine. In such proposal, latency sensitive applications report to mechanism whenever they are under interference. Then, the mechanism uses information collected in that instant to predict when latency application will suffer interference again. From this prediction, the mechanism throttles the batch application before it imposes interference to latency application one more time. Thus, that work just proposed a way to work around the interference suffered by a specific class of application by monitoring the conditions that lead to occurrence of the interference. So, the root of interference problem was not investigated and, as consequence, a solution to determine interference suffered by any set of applications was not proposed.  

Other works, however, investigated the cross-application interference problem in a broader sense and proposed solutions to determine or alleviate interference experienced by co-located applications in general. 

Mury \textit{et al.} \cite{mury2014concurrency} argued that cross-application interference could be determined by adopting a classification of applications. Such qualitative approach is based on the \textit{Thirteen Dwarfs} which classifies applications according to computational methods usually adopted in scientific computing. Such classification is not suitable to determine interference because our experiments showed that two applications belonging to Dense Linear Algebra class presented distinct interference levels. Besides that, our results also pointed out that a same application, namely PTRANS, can present distinct interference levels when solving  instances with different sizes. Actually, experiments conducted in \cite{gupta2013hpc} also showed this same behavior when  applications EP and LU presented distinct interference levels when solving also instances of different sizes. 

Jin \textit{et al.} \cite{jin2015ccap}  classified applications in three SLLC access classes called (i) cache-pollution, (ii) cache-sensitive and (iii) cache-friendly. These classes were further used to propose a VMP strategy with goal to alleviate interference by co-locating applications with compatible SLLC access profiles. That work claimed that cache-pollution applications should be preferably co-located with cache-friendly applications rather than  being co-located with cache-sensitive ones. However, through some practical  experiments they showed that approach may fail.  More specifically, EP and IS,  classified as cache-friendly, suffered distinct interference levels when co-located with CG,  categorized as a cache-pollution application. Besides that, although both CG and FT were classified as cache-pollution applications, CG did not present interference when co-located with itself, while FT suffered an interference when co-located with CG. These results  show that a  qualitative approach based on SLLC access pattern is not suitable to determine  cross-application interference precisely.

Unlike the aforementioned works, Gupta \textit{et al.} \cite{gupta2013hpc} adopted a quantitative approach to investigate cross-application interference. They studied the relation between interference and the number of SLLC references per second performed simultaneously by co-located applications. As a result, they identified a maximum access limit that ensures a free interference co-location. Although this work proposed a quantitative strategy to explain interference problem, only one shared resource,  SLLC, was considered. Our experiments showed that others shared and non-sliceable resources such as virtual network can also influence interference and, consequently, should be systematically evaluated.

Moreover, two of the previously described works,  by Gupta \textit{et al.} \cite{gupta2013hpc} and  by Jin \textit{et al.} \cite{jin2015ccap}, just inform whether applications should be or not co-located, without quantifying the  level of interference suffered by them. However, the information about interference level can enrich the VMP process, specially in cases where only applications presenting high cross-interference levels are available.

To the best of our knowledge, this paper is the first one that proposes a quantitative model for predicting  cross-application interference level by considering  accesses to SLLC, DRAM and virtual network, jointly. We claim  that a  quantitative approach, that considers the amount of accesses of co-located applications to shared resources simultaneously, is more suitable to  determine cross-application interferences.

\section{Generating a Cross-application Interference Dataset}

In order to create an interference dataset, several  co-locating synthetic applications with distinct access levels were executed and evaluated. 

Those synthetic applications  with distinct access patterns to each shared resource were obtained from a template  with  different input parameters. 

Next, some preliminary concepts are presented in subsection \ref{sec:concepts}. Then, the proposed  application template is described in subsection \ref{sec:synth_model}. In subsection \ref{sec:synthetic} we present the used  synthetic applications and in subsection \ref{sec:results} results of  concurrent executions of  these synthetic applications are presented. 

\subsection{Preliminary Concepts}
\label{sec:concepts}

 We present here two fundamental definitions about cross-application interference used along this paper: the \textit{accumulated access} to shared resources and the \textit{interference level} suffered by co-located applications.

As defined in Equation \ref{eq:access}, the access of an application $i$ to a shared resource $s$ is equal to the sum of access of all virtual machines holding this application to this shared resource, where $nV_i$ is the total number of virtual machines hosting $i$ and $V_{i,j,s}$ is the amount of access of virtual machine $V_j$  to a  shared resource $s$. In this work the amount of access to  SLLC and DRAM are measured in terms of millions of references per second (MR/s), while the access to virtual network is expressed as the amount of megabytes transmitted per second (MB/s). 

\begin{equation}
    \label{eq:access}
    A_{i,s} = \sum_{j=1}^{nV_{i}} V_{i,j,s}
\end{equation}

From Equation \ref{eq:access}, we defined the \textit{accumulated access} of all applications  to a shared resource. Thus, as defined in Equation \ref{eq:accumulated_access}, this accumulated access to a shared resource $s$ is equal to the sum of access of all applications co-located in same physical machine, where $nA$ is the total number of applications co-located in a same physical machine and $A_{i,s}$ is the amount of access of application $A_{i}$ to shared resource $s$. 
This accumulated access represents the total amount of pressure to a shared resource in a given time interval.

\begin{equation}
    \label{eq:accumulated_access}
    T_{s} = \sum_{i=1}^{nA} A_{i,s}
\end{equation}

We argue that the amount of accumulated access to specific shared resources is directly related to the  slowdown  suffered by co-located applications. In this work,  slowdown   is defined as the ratio of the execution time achieved by application when executed concurrently with co-located applications ($C_{i,X}$) to the one achieved from isolated execution ($T_{i}$) less 1, where $X$ is the set of applications co-located in a same physical machine, as shown in  Equation \ref{eq:slowdown},.

\begin{equation}
    \label{eq:slowdown}
    S_{i,X} = \frac{C_{i,X}}{T_{i}} - 1
\end{equation}

Then, the \textit{cross-application interference level} is defined  as the average  slowdown of  applications  when co-located with other ones in a same physical machine.

For example, suppose that the execution times of two applications, namely A and B,  in a dedicated processor were equal to 60 and 80 seconds. Suppose also that these both applications, when concurrently executed in that processor, spent 100 seconds. In such scenario, the slowdown of applications A and B would be, respectively, 67\% and 25\%, which represents how much additional time  these applications needed to complete their executions when co-located in a same processor. 
The cross-application interference level between applications A and B would be 46\%. Thus, applications A and B, would suffer, in average, 46\% of mutual interference when co-located in a same processor.

\subsection{Generating Synthetic Applications}
\label{sec:synth_model}

The access contention  to shared resources  is the  main cause of the interference suffered by applications co-located  in a same physical machine.    
To study cross-application interference, most of related works\ref{sec:related} employs  real applications provided by  traditional HPC benchmarks. 
However,   a real application does not allow to control the  number of accesses to each shared resource,  and consequently, to evaluate systematically the  relation between concurrent accesses and  the resulting interference.
Thus, we propose  an  \textit{application template} from which  synthetic  applications with distinct access levels are created. From this template,  we can  create an  application which performs a high access pressure to SLLC, while keeping a low access level to virtual network, for example. Thus, a set of synthetic applications created from that template allows to observe interference  before different levels of accesses  to SLLC, DRAM and virtual network. 

In order to represent the  usual behavior  of  HPC applications \cite{silva2015memory},  the proposed synthetic application template  presents, alternately, two distinct and well-defined phases. The first one, called \textit{Computation Phase}, represents the phase at  which  the application performs tasks involving calculation and data movement. The other one, namely \textit{Communication Phase}, is  the phase where  the application exchanges information among computing pairs.  

The proposed  synthetic application template is shown in Algorithm \ref{alg:synth}. Firstly,   the synthetic application  executes the \textit{Main Loop} (lines 1 to 13) whose total number of iterations is controlled by parameter $\omega$. This parameter  leverages the total execution time of application.  Next, the synthetic application executes \textit{Computation Phase Loop} (lines 2 to 9)  at which  it performs the \textit{Computation Phase} with the number of iterations defined by $\alpha$. This \textit{Computation Phase} is based on  the benchmark STREAM which is widely used to measure performance of memory subsystems \cite{papagiannis2014hybrid} \cite{xavier2013performance}. In order to measure sustainable memory bandwidth, this benchmark executes  four simple vector kernels, as  presented in Table \ref{tab:stream}.  Because SUM presents the highest tax of memory access, it was chosen to be included in the proposed template.

    \begin{algorithm}[h]
			\caption{Synthetic application template}
			\label{alg:synth}
            \begin{algorithmic}[1]
            \Statex \textbf{Input parameters:} $\omega$,$\alpha$, $\gamma$, $\delta$, $\theta$, $\beta$, $\lambda$
            \Statex \textit{/* Main Loop*/}
            \For{x=1 to $\omega$}
                \Statex \ \ \ \ \ \ \textit{/* Computation Phase Loop*/}        
            	\For{y=1 to $\alpha$}
					\Statex \ \ \ \ \ \ \ \ \ \ \ \ \textit{/* Memory Access Loop*/} 
					\For{i=1 to $\gamma$ \textbf{step} $\delta$}
					    \State A[i] = B[i] + C[i];
                        \Statex \ \ \ \ \ \ \ \ \ \ \ \ \ \ \ \ \ \  \ \textit{/*Compute-intensive Loop*/}					  
                        \For{k=1 to $\theta$}
                            \State T = SquareRoot(T);
                        \EndFor
					\EndFor
				\EndFor
				\Statex \ \ \ \ \ \ \textit{/* Communication Phase Loop*/}
                \For{z=1 to $\beta$}
				    \State All-to-All-Communication(D,$\lambda$);
			    \EndFor

			\EndFor
			\end{algorithmic}
        \end{algorithm}

The SUM operation is executed by  the inner loop \textit{Memory Access Loop} (lines 3 to 8) which is controlled by two input parameters, $\gamma$ and $\delta$. The first one defines the  sizes of vectors $A$, $B$ and $C$, and is indirectly used to determine application's Working Set Size (WSS) \cite{lelli2012experimental} \cite{gupta2013hpc}. A small WSS usually increases application's cache hit ratio because all data needed by it in a given time interval can be entirely loaded in cache. On the other hand, when the  application has a WSS greater than cache capacity, its cache hit ratio decreases because all data is fetched from main memory. Thus, WSS can be used to control cache hit ratio  and, consequently,  control the number of DRAM references per second also.

\begin{table}[h]
    \centering
    
    \label{tab:stream}
    \begin{tabular}{|c|c|c|}
        \hline
        \textbf{Name} & \textbf{Operation} & \textbf{Bytes per Iteration}\\ \hline
        COPY           &  \texttt{a[i] = b[i]} &16                   \\ \hline
        SCALE          &  \texttt{a[i] = b[i]*q} &16                   \\ \hline
        SUM           &  \texttt{a[i] = b[i] + c[i]} &24                   \\ \hline
        TRIAD           &  \texttt{a[i] = b[i] + c[i]*q} &24                   \\ \hline
    \end{tabular}
    \caption{STREAM kernels}
\end{table}

The second parameter of \textit{Memory Access Loop}, namely $\delta$,  controls the step at which the vector elements are accessed. Thus, when $\delta$ is equal to 1, all elements of vectors $A$, $B$ and $C$ are accessed consecutively, resulting in a high cache hit ratio. Otherwise, when $\delta$ is set to a high value, more   data are fetched from   main memory,  resulting in   performance degradation.  In other words, this parameter provides another way to control cache hit ratio  and  to manipulate the number of DRAM references per second. 

Thus, the number of DRAM references per second   and application's cache hit ratio can be controlled by performing a fine tuning of both parameters $\gamma$ and $\delta$. When the  application presents a high cache hit ratio, DRAM receives few  references per second because data is already available in SLLC. Likewise, the number of memory references increases when application presents a low cache hit ratio. 

Besides controlling DRAM access, these both parameters allows to handle the number of SLLC references per second as well. When the  application presents a low cache hit ratio, the number of SLLC references per second decreases. This happens because, before accessing  new data,  the previous referenced data,  not found in SLLC,  has to be fetched from DRAM. Consequently, the rate at which application performs data accesses is reduced, decreasing also the number of SLLC references per second. On the other hand,  when the  application presents a high cache hit ratio, the number of SLLC references per second increases.  Because most data is rapidly fetched from SLLC, more memory accesses can be performed per second. 

After performing the SUM operation (line 4), synthetic application executes \textit{Compute-intensive Loop} which repeatedly calculates the square root of variable $T$ (line 6). This loop, whose number of iterations is defined by $\theta$, allows to make application more or less compute-intensive. Note that when $\theta$ is set to a high value, the number of memory references  decreases. This happens because variable $T$, being frequently referenced,  is kept stored  in  the first cache level (cache L1), preventing memory subsystem lower levels of  being accessed. As a result, both number of SLLC and DRAM references per second decreases. Thus, together with $\gamma$ and $\delta$, $\theta$ is also used to manipulate the number of DRAM and SLLC references per second.  

After \textit{Computing Phase Loop} execution, synthetic application performs \textit{Communication Phase} by executing \textit{Communication Phase Loop} (lines 10 to 12) whose number of iterations is determined by input parameter $\beta$. This phase is based on MP\_Bench benchmark  \cite{filgueira2012dynamic}.  
From all MPI operations executed by this benchmark,  \texttt{MPI\_Alltoall} was particularly interesting for our work  because it is widely used in scientific applications. When using this collective operation, all application's processes send and receive to/from each other the same  amount of  data \cite{steffenel2007assessing} \cite{hochstein2008pilot} \cite{li2011optimizing}. 

For each \textit{Comunnication Phase Loop} iteration, the application executes \textit{All-to-All-Communication()} function (line 11)  that employs \texttt{MPI\_Alltoall}  of a  vector $D$ whose size is defined by the input parameter $\lambda$. So, this input parameter is used to handle the number of bytes transmitted in the  virtual network.

Synthetic applications with distinct access profiles can be generated by varying properly all of these aforementioned input parameters, whose descriptions are summarized as next.

\begin{itemize}
    \item $\omega$: application's total number of iterations.  
    \item   $\alpha$: total number of iterations of Computation Phase.
    \item   $\beta$: total number of iterations of Communication Phase.  
    \item  $\gamma$: Working Set Size (WSS).          
    \item   $\delta$: step at which the vector elements in \textit{Memory Access Loop} are accessed.
    \item    $\theta$: level of compute-intensiveness.            
    \item    $\lambda$: amount of data exchanged among processes.        
\end{itemize}

Unlike adopting real applications,  with this proposed synthetic application template,  several  applications that perform  distinct access pressure to SLLC, DRAM and virtual network can be generated,  providing a proper way to investigate systematically  the relation between number of accesses and cross-application interference.

\subsubsection{Used Synthetic Applications}
\label{sec:synthetic}

In this subsection, we present the set of synthetic applications generated from  the previously presented  application template. Applications with distinct access rates were generated, considering three target access levels for each  of the three shared resources. The  access rates to each shared resource are expressed by distinct metrics, such as  number of references  to memory per second or  transmitted  bytes per second, and the range of  those values are  also different.  To treat those access rates jointly, we normalized those values in  an interval between 0.0 and 1.0, where score 1.0 represents the highest possible  access rates achieved  by an  application  based on the proposed template, and score 0.0 represents no access. These scores, in our work,   represent  different   access levels to the shared resources.  Then, we created applications with \textit{high}, \textit{medium} and \textit{low} access levels  to SLLC, DRAM and virtual network, where the high access level corresponds, in our proposal, to the highest access rate to each shared resource, and medium and low access levels correspond to 50\% and 10\% of this high access rate, respectively. 

To generate those applications with these distinct access levels, we varied input parameters of application template and monitored resulted access rate to each shared resource by using monitoring tools such as PAPI (Performance Application Programming Interface) \cite{chiang2016kernel}, OProfile \cite{tsao2012seprof} and SAR (System Activity Report) \cite{popiolek2013monitoring}

\begin{figure}[h] 
    \centering
    \includegraphics[scale=0.5]{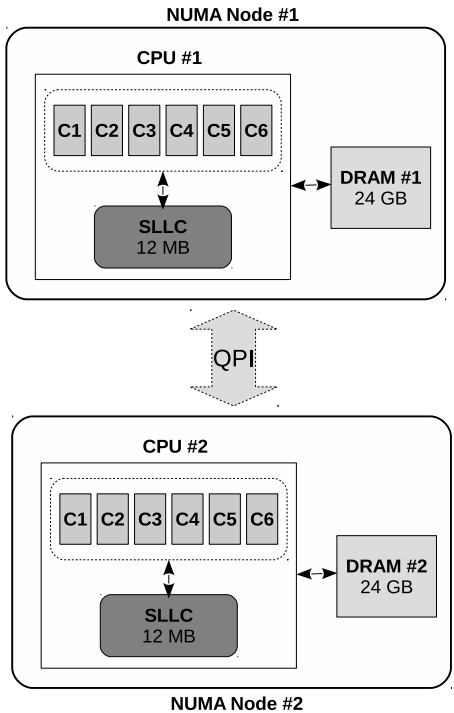}
    \caption{NUMA nodes and processors used in our experiments}
    \label{fig:itautec}
\end{figure}

We executed this set of synthetic applications in  a Itautec MX214 server whose configuration details are described in Table \ref{tab:mx214}. As  illustrated in Figure \ref{fig:itautec}, this server is equipped with two NUMA nodes interconnected by a QPI (Quick Path Interconnect) of 6.4 GT/s, where each NUMA node has 24GB of DRAM memory and is endowed with a Intel Xeon X5675 3.07GHz processor. Each processor has six  cores that share a 12MB SLLC unit. Moreover, the virtual environment was provided by KVM hypervisor running on top of Ubuntu Server.  


\begin{table}[h]
\centering
\label{tab:mx214}
\begin{tabular}{|l|l|}
\hline
\textbf{Model}              & Itautec MX214                                     \\ \hline
\textbf{CPU}                & \multicolumn{1}{c|}{2x Intel Xeon X5675 3.07 GHz} \\ \hline
\textbf{DRAM}               & 48 GB DDR3 1333 MHz                               \\ \hline
\textbf{Disk}               & 5.8 TB SATA 3 GB/s                              \\ \hline
\textbf{QPI}               & 6,4 GT/s
\\ \hline

\textbf{S.O.}               & Ubuntu 15.04                                      \\ \hline
\textbf{Kernel}             & 3.19.0-15                                         \\ \hline
\textbf{Hypervisor}         & KVM                                               \\ \hline
\textbf{Hardware Emulation} & Qemu 2.2.0                                        \\ \hline
\end{tabular}
\caption{Configuration of server used in experiments}
\end{table}

In our proposal, each application uses six virtual machines,  each one deployed in  one core, allocating 4GB of main memory.  We used  CPU affinity to deploy half of the virtual machines of the application in each NUMA node, i.e., three  virtual machines were deployed in ``NUMA Node \#1", while the others were deployed in ``NUMA Node \#2", in a dedicated machine. In this scenario, the access of synthetic application to shared resources, specifically SLLC and DRAM, is balanced over two NUMA nodes, avoiding self-interference. 
Moreover, that configuration will be helpful to evaluate cross-application interference as  discussed in the next section.

The set of the generated synthetic applications and the corresponding  execution profiles are described in Table \ref{tab:synthetic}. All applications execute the same number of iterations ($\omega$ = 25) and parameters $\alpha$ and $\omega$ were set to ensure that they spent approximately  the same amount of time executing \textit{Computation} and \textit{Communication Phases}. Moreover,  all scores were rounded to one decimal place. This explains, for example, why applications S1 and S7, although have presented distinct absolute values, were classified in the same SLLC score. 

Applications S1, S7 and S13 achieved the highest SLLC number of references per second. In order to reach this high SLLC access, we adjusted  the input parameters  to ensure that all memory references were directly satisfied by SLLC, what resulted in a 0.0 DRAM score. Thus, although score 0.0 was not considered as one of the three target access levels, there is no way to achieve the highest number of SLLC references per second without reducing drastically the number of accesses to DRAM. 

On the other hand, the number of memory references satisfied by SLLC has to decrease  to rise the number of DRAM references per second. To achieve a high DRAM access, all memory references should result in  accesses to   main memory, i.e.,  the SLLC hit ratio  must be close to 0\%. However, even in that case, the number of SLLC references per second  is not equal to zero, because all references to DRAM are also treated by SLLC. This explains why applications S4, S10 and S16, which achieved the highest number of DRAM references per second, exhibited a SLLC score equal to 0.3. 

Thus, concerning the memory subsystem, we were not able not generate all possible combinations involving the three access levels.  A high number of accesses to SLLC  implicates in a low number of accesses to DRAM. As a consequence,it is not possible to generate an application which both SLLC and DRAM scores  equal to 1.0 or an application which performs, simultaneously, a high and medium access level to SLLC and DRAM, for example. 

Besides that, that behavior also resulted in some unexpected combinations as the one presented by application S4, that presented DRAM and SLLC scores  equal to 1.0 and 0.3, respectively, though this last score was not considered as one of the target access levels. 

At last, concerning virtual network, the highest amount of transmitted bytes was achieved by increasing input parameter $\lambda$ up to  reaching the maximum amount of data that the  hypervisor is able to handle at same time. We varied $\lambda$  to find out the virtual network saturation threshold. When this limit is exceeded, the amount of bytes transmitted per second  decreases, regardless of the increasing of $\lambda$.

Although we did not generate all possible combinations of applications, we were able to create a synthetic workload with distinct computational burden, suitable for conducting a deep evaluation of the cross-application interference problem.  

\begin{table*}[h]
\centering

\label{tab:synthetic}

\begin{tabular}{c|c|c|c|c|c|c|c|c|c|c|c|c|c|}
\cline{2-14}
\textbf{}                                  & \multicolumn{3}{c|}{\textbf{Absolute Value}}     & \multicolumn{3}{c|}{\textbf{Score}}          & \multicolumn{7}{c|}{\textbf{Parameters}}                                                                                                  \\ \hline
\multicolumn{1}{|c|}{\textbf{Application}} & \textbf{SLLC} & \textbf{DRAM} & \textbf{INN} & \textbf{SLLC} & \textbf{DRAM} & \textbf{INN} & \textbf{$\omega$} & \textbf{$\alpha$} & \textbf{$\beta$} & \textbf{$\gamma$} & \textbf{$\delta$} & \textbf{$\theta$} & \textbf{$\lambda$} \\ \hline
\multicolumn{1}{|c|}{S1}                   & 1635          & 4             & 300          & 1.0          & 0.0          & 0.1         & 25                & 120000            & 5200             & 7000              & 512               & 0                 & 22600              \\ \hline
\multicolumn{1}{|c|}{S2}                   & 851           & 61            & 324          & 0.5          & 0.1          & 0.1         & 25                & 90000             & 5200             & 9000              & 1024              & 6                 & 22600              \\ \hline
\multicolumn{1}{|c|}{S3}                   & 239           & 41            & 312          & 0.1          & 0.1          & 0.1         & 25                & 40000             & 5200             & 11500             & 2048              & 22                & 22600              \\ \hline
\multicolumn{1}{|c|}{S4}                   & 444           & 444           & 318          & 0.3          & 1.0           & 0.1         & 25                & 7500              & 5200             & 30000             & 512               & 0                 & 22600              \\ \hline
\multicolumn{1}{|c|}{S5}                   & 224           & 224           & 324          & 0.1          & 0.5          & 0.1         & 25                & 2700              & 5200             & 39000             & 512               & 21                & 22600              \\ \hline
\multicolumn{1}{|c|}{S6}                   & 797           & 240           & 318          & 0.5          & 0.5          & 0.1         & 25                & 20000             & 5200             & 11800             & 256               & 2                 & 22600              \\ \hline
\multicolumn{1}{|c|}{S7}                   & 1597          & 18            & 2892         & 1.0          & 0.0          & 1.0         & 25                & 120000            & 1500             & 7000              & 512               & 0                 & 749568             \\ \hline
\multicolumn{1}{|c|}{S8}                   & 890           & 43            & 2810         & 0.5          & 0.1          & 1.0         & 25                & 90000             & 1500             & 9000              & 1024              & 6                 & 749568             \\ \hline
\multicolumn{1}{|c|}{S9}                   & 220           & 49            & 2910         & 0.1          & 0.1          & 1.0         & 25                & 40000             & 1500             & 11500             & 2048              & 22                & 749568             \\ \hline
\multicolumn{1}{|c|}{S10}                  & 438           & 438           & 2832         & 0.3          & 1.0          & 1.0         & 25                & 7500              & 1500             & 30000             & 512               & 0                 & 749568             \\ \hline
\multicolumn{1}{|c|}{S11}                  & 214           & 214           & 2892         & 0.1          & 0.5          & 1.0         & 25                & 2700              & 1500             & 39000             & 512               & 21                & 749568             \\ \hline
\multicolumn{1}{|c|}{S12}                  & 817           & 241           & 2838         & 0.5          & 0.5          & 1.0         & 25                & 20000             & 1500             & 11800             & 256               & 2                 & 749568             \\ \hline
\multicolumn{1}{|c|}{S13}                  & 1575         & 22            & 1392         & 1.0          & 0.0          & 0.5         & 25                & 120000            & 150000           & 7000              & 512               & 0                 & 150000             \\ \hline
\multicolumn{1}{|c|}{S14}                  & 890           & 52            & 1362         & 0.5          & 0.1          & 0.5         & 25                & 90000             & 150000           & 9000              & 1024              & 6                 & 150000             \\ \hline
\multicolumn{1}{|c|}{S15}                  & 228           & 49            & 1335         & 0.1          & 0.1          & 0.5         & 25                & 40000             & 150000           & 11500             & 2048              & 22                & 150000             \\ \hline
\multicolumn{1}{|c|}{S16}                  & 438           & 438           & 1375         & 0.3          & 1.0          & 0.5         & 25                & 7500              & 150000           & 30000             & 512               & 0                 & 150000             \\ \hline
\multicolumn{1}{|c|}{S17}                  & 221           & 221           & 1404         & 0.1          & 0.5          & 0.5         & 25                & 2700              & 150000           & 39000             & 512               & 21                & 150000             \\ \hline
\multicolumn{1}{|c|}{S18}                  & 824           & 239           & 1380         & 0.5          & 0.5          & 0.5         & 25                & 20000             & 150000           & 11800             & 256               & 2                 & 150000             \\ \hline
\end{tabular}
\caption{Generated synthetic applications and the corresponding  execution profiles when executed in a dedicated machine}
\end{table*}

\subsection{Measuring Cross-applications Interference}

\label{sec:results}

In this section, we present  experiments  to determine  cross-applications interference.  The previously presented synthetic applications were executed in a two-by-two fashion  to obtain the resulting interference level in several cases. 
Because each synthetic application used half of available resources (memory and CPU),  we were able to co-locate two of those applications in the physical machine,  without exceeding the available resources in the system.  That full allocation represents a realistic scenario, usually found in clouds environments, where all resources available in a physical machine are fully allocated to maximize resource utilization \cite{pires2015virtual} \cite{shirvani2016server}.

\begin{figure}[h] 
    \centering
    \includegraphics[scale=0.47]{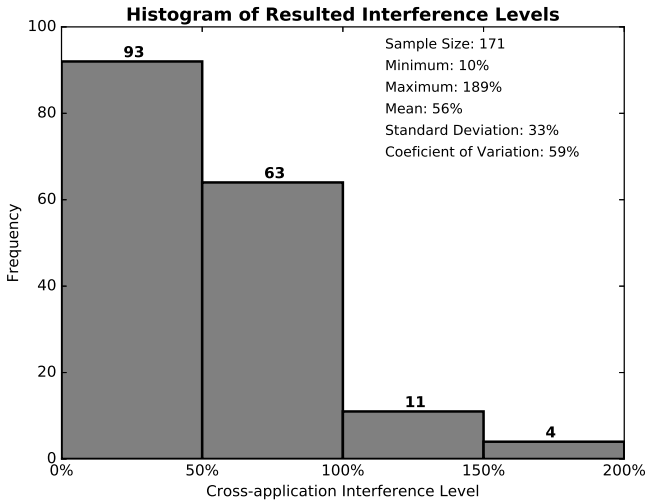}
    \caption{Histogram of resulted interference levels achieved from synthetic experiments}
    \label{fig:interf_results}
\end{figure}

 The generated synthetic applications do not present exactly the same execution time,  so to  keep concurrency among co-located applications until the end of the experiment, the smaller execution  time application  was re-started  automatically as many times as necessary to cover the  entire execution of the longer application. We adopted such approach  to fairly measure the interference suffered by both applications, regardless their  execution times.  

As the synthetic workload is composed of 18 applications, our interference experiments comprised 171 concurrent executions whose results are summarized in Figure \ref{fig:interf_results}.  Those results show that cross-applications interference can vary drastically,  from 10\% to 189\%, depending on which applications are  co-located in the same physical machine. As can be seen  in Figure \ref{fig:interf_results}, around 54\% of the total co-executions (93 occurrences) achieved an interference level less than 50\%, while 37\% of co-locations (63 occurrences) suffered interference levels between 50\% and 100\%. Besides that, in around 9\% of all cases (15 occurrences) co-location applications reached interference levels greater than 100\%. These results presented a coefficient of variation\footnote{Also known as relative standard deviation, the coefficient of variation is defined as the ratio of the standard deviation to the mean.} close to 56\% which allows to assert that these synthetic experiments comprised a large range of interference levels. 

\begin{figure}[h] 
    \centering
    \includegraphics[scale=0.47]{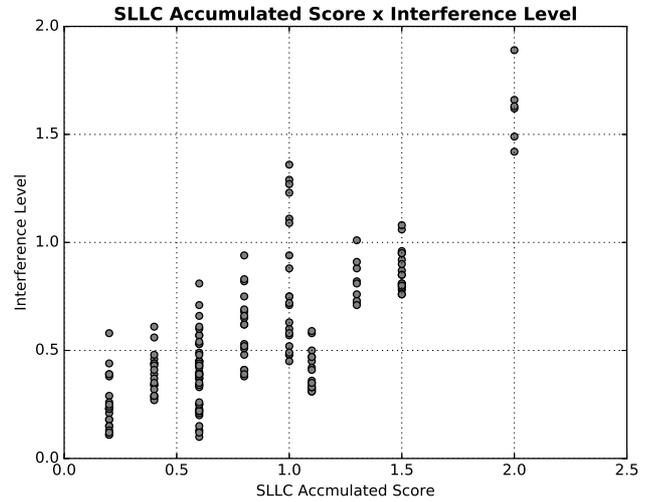}
    \caption{Scatter plot of SLLC accumulated score against interference level}
    \label{fig:sllcxinterf}
\end{figure}

An initial analysis accomplished in these results revealed that there is a correlation between interference level and SLLC accumulated score. As can be seen in Figure \ref{fig:sllcxinterf}, interference level tends to increase as SLLC accumulated score rises. Indeed, the Pearson's correlation coefficient between SLLC accumulated score and interference level is around 0.76, indicating a strong, positive and linear relationship between these both variables. Thus, this observation corroborates the hypothesis that the amount of access performed in shared resources can really influence the interference  suffered by co-located applications.

In addition, these experiments allowed to confirm that mutual access performed in other shared resources besides SLLC can also impact the interference level. Consider, for example,  SLLC accumulated score  equal to 0.40,  in Figure \ref{fig:sllcxinterf}, it may occur  in  cases with  distinct interference levels. In other words, although SLLC access presents a strong correlation with interference level, there is another factor influencing it.

\begin{figure}[h] 
    \centering
    \includegraphics[scale=0.47]{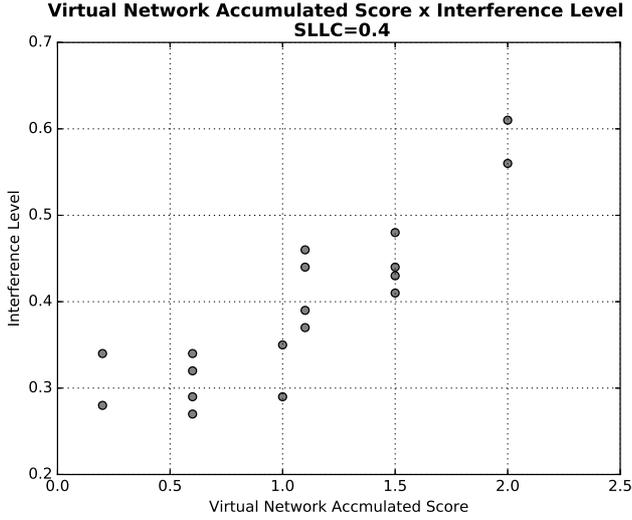}
    \caption{Scatter plot of virtual network accumulated score against interference level when SLLC was equal to 0.40}
    \label{fig:netxinterf}
\end{figure}

 Actually, the interference level increases as virtual network also does. As illustrated in Figure \ref{fig:netxinterf}, when virtual network accumulated score is equal to 0.2 and 2.0, the corresponding interference levels are around 28\% and 60\%, respectively. For the same SLLC  accumulated score, the interference level suffered by co-located applications varies more than 30\% depending on the amount of access to virtual network.

Moreover,  some co-locations, even though performed almost the  same amount of accumulated access to all  shared resources,  present interference levels that varied in more than 45\%. In the subset of interference results, listed in Table \ref{tab:concurrency}, for example,  the co-location ``S15xS7'' suffered an interference level 46\% higher than the  co-location ``S14xS8'', although both have the same virtual network accumulated score,  1.5, and differ slightly about DRAM and SLLC accumulated scores. Applications S15 and S7 present, individually,  distinct SLLC access values, the former  performs much less access  to   SLLC than the latter.  This explains why this co-location does not present a high interference level, even achieving a high SLLC accumulated access. On the other hand,  S15 and S7,  although present a similar  SLLC accumulated access,   evenly compete  for  SLLC,  resulting in a high interference level.  As can be seen in Table \ref{tab:concurrency}, this also happens in  co-locations that present virtual network accumulated scores equal to 0.6 and 0.2. 

\begin{table}[h]
\centering

\label{tab:concurrency}
\begin{tabular}{|c|c|c|c|c|}
\hline
\multirow{2}{*}{\textbf{Co-execution}} & \multicolumn{3}{c|}{\textbf{Accumulated Score}} & \multirow{2}{*}{\textbf{\begin{tabular}[c]{@{}c@{}}Interference \\ Level\end{tabular}}} \\ \cline{2-4}
                                       & \textbf{SLLC}  & \textbf{DRAM}  & \textbf{NET}  &                                                                                         \\ \hline
S1 x S3                                & 1.1            & 0.1            & 0.2           & 34,10\%                                                                                 \\ \hline
S2 x S2                                & 1.0            & 0.2            & 0.2           & 71,12\%                                                                                 \\ \hline
S13 x S3                               & 1.1            & 0.1            & 0.6           & 31,51\%                                                                                 \\ \hline
S14 x S2                               & 1.0            & 0.2            & 0.6           & 71,83\%                                                                                 \\ \hline
S15 x S7                               & 1.0            & 0.2            & 1.5           & 41,67\%                                                   
                              \\ \hline
S14 x S8                               & 1.1            & 0.1            & 1.5           & 87,97\%                                                                                 \\ \hline

\end{tabular}
\caption{Subset of synthetic interference experiments}
\end{table}

So, besides accumulated access performed to shared resources, the similarity of application's access burden has a direct impact in interference level suffered by applications when co-located in a same  machine. Indeed, this justifies the difference between interference levels achieved by co-locations listed in Table \ref{tab:concurrency}. 

In order to measure the level of similarity between two applications, we define in Equation \ref{eq:similarity_factor} the \textit{similarity factor}. The \textit{similarity factor} of two applications regarding to a shared resource $s$ is calculated as the difference between 1 and the absolute value resultant from the difference between the amount of individual access that applications $i$ and $j$ perform in a shared resource $s$. From Equation \ref{eq:similarity_factor}, we defined the \textit{global similarity factor} as being the average of all similarity factors calculated for each pair of applications co-located in a physical host.

\begin{equation}
    \label{eq:similarity_factor}
    F_{i,j,s} = 1 - |A_{i,s} - A_{j,s}| 
\end{equation}

From results reached on these experiments, we can draw some preliminary conclusions about cross-application interference. 

\begin{itemize}
    \item There is a strong correlation between amount of simultaneous accesses to SLLC and level of interference suffered by applications. 
    \item Although SLLC access rate presents a strong correlation with interference level, the simultaneous access to other shared resources, such as virtual network, can contribute to increase interference as well.  
    \item Besides total amount of access to shared resources, the similarity between application's access burden can also impact interference level suffered by co-located applications.   
\end{itemize}


That scenario  justifies the adoption  of a quantitative approach for predicting the interference level suffered by co-located applications.  As shown,  the cross-application interference  level  is  influenced by the collective effect of more than one variable, what ...pareaqui   treated as a multivariate problem

\section{Quantitative Cross-application Interference Prediction Model}

In this section, we describe process accomplished for building our proposed quantitative prediction model by using Multiple Regression Analysis. First, in subsection \ref{sec:mra}, we briefly introduce Multiple Regression Analysis, while in subsection \ref{sec:building} we describe how this technique was employed to build our prediction model. 

\subsection{A Brief Introduction to Multiple Regression Analysis}
\label{sec:mra}

Multiple Regression Analysis (MRA) is a multivariate statistical technique that allows to explain the relationship between one dependent variable and, at least, two independent variables. This technique is used to create a model, a.k.a. statistic variable, able to predict the value of the dependent variable from the known values of independent variables \cite{ngo2012steps}. 

\begin{figure}[h] 
    \centering
    \includegraphics[scale=0.47]{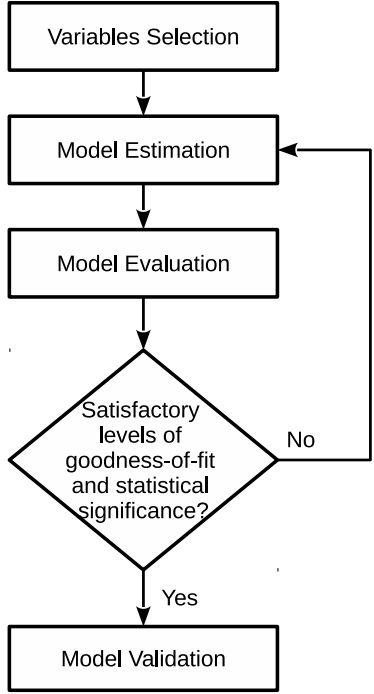}
    \caption{Model building flowchart}
    \label{fig:regression_steps}
\end{figure}

Basically, MRA is comprised of four macro steps as illustrated in flowchart of Figure \ref{fig:regression_steps}. Firstly, in \textit{Variables Selection} step, researcher selects the most likely variables to explain the behavior of the response variable. Although some statistical techniques such as matrix of correlation can provide some insights about what variables must be chosen, this process is mainly guided by the researcher's knowledge about the problem \cite{hair2006multivariate} \cite{ngo2012steps}. 

After that, researcher executes the \textit{Model Estimation} step, where the terms of the model are determined and their coefficients are automatically estimated by using the Least Squares Method. Next, researcher proceeds to the \textit{Model Evaluation} step in order to evaluate goodness-of-fit level, i.e., how satisfactorily the estimated model fits to data used in building process. Besides that, researcher also assesses statistical significance of regression and coefficients. In case that estimated model does not present a satisfactory goodness-of-fit level or a desired statistical significance level, a new model should be estimated by executing, again, the \textit{Model Estimation} step. Otherwise, estimated model is ready to be validated in \textit{Model Validation} step by using an ``unseen" dataset, i.e, a dataset not used previously during the process of model estimation \cite{hair2006multivariate} \cite{ngo2012steps}  \cite{mason1991collinearity} \cite{vandekerckhove2014model}. 

Although non specialized programs such as Excel and Matlab can be used for building a model from MRA, this process is usually accomplished by using commercial statistical packages such as Eviews \cite{min2011eviews}, SAS \cite{agresti2011categorical}, and Minitab \cite{wallshein2015software}, or free software ones like Gretl \cite{baiocchi2003gretl} or R \cite{r2016}. Such specific tools provides a full multivariate analysis toolbox that allows to perform a deep analysis on estimated model  \cite{wijayasiriwardhane2010component} \cite{nassif2013towards} \cite{montgomery2015introduction}.

\subsection{Model Building}
\label{sec:building}

By using Minitab version 17.1.0, we followed the aforementioned process to build our proposed model. At first, we executed the \textit{Variables Selection} step in order to identify what variables should be selected as independent ones. As synthetic dataset was generated specifically to investigate interference, this selection process was straightforward. Therefore, we determined accumulated scores and global similarity factors to all of the three shared resources as independent variables and cross-application interference level as the dependent one. So, we expect to generate a model able to predict the interference level from the known values of accumulated scores and global similarity factors. 

After \textit{Variable Selection} step, we repeatedly executed both \textit{Model Estimation} and \textit{Model Evaluation} steps till reach a parsimonious model with satisfactory levels of goodness-of-fit and statistical significance. A parsimonious model is able to perform better out-of-sample predictions because, due to its simplicity, it is not usually overfitted to the sample \cite{vandekerckhove2014model}. 

Our quantitative model for predicting interference is described in Equation \ref{eq:model} whose terms are listed in Equations \ref{eq:t1}, \ref{eq:t2} and \ref{eq:t3}. As can be seen in estimated model, global similarity factors, namely $G_{sllc}$, $G_{dram}$ and $G_{net}$, were employed to weigh the influence that accumulated accesses, namely $T_{sllc}$, $T_{dram}$ and $T_{net}$, impose in interference. Moreover, the total amount of access performed in DRAM was ponder by SLLC accumulated access since all access requests to DRAM are firstly treated in SLLC. At last, it is worth mentioning that the constant as well as quadratic terms were not included in model in order to make it as simple as possible.


\begin{equation}
    \label{eq:model}
    I = 0.7498*T1 + 0.1598*T2 + 0.1456*T3
\end{equation}

Where,

\begin{equation}
    \label{eq:t1}
    T1 = T_{sllc} * G_{sllc}
\end{equation}

\begin{equation}
    \label{eq:t2}
    T2 = T_{net} * G_{net}
\end{equation}

\begin{equation}
    \label{eq:t3}
    T3 = T_{dram} * T_{sllc}* G_{sllc} 
\end{equation}

From coefficients estimated by each term, we can state that the simultaneous and accumulated access to SLLC really imposes the higher influence in interference. However, as previously discussed, the access to other shared resources can also affect interference since terms regarding to virtual network and DRAM, respectively T2 and T3, presented, together, a coefficient very close to 0.31. Thus, depending on the value of SLLC accumulated score, the interference will be primarily determined by the access performed to DRAM and virtual network.   

This model presented an Adjusted Coefficient of Regression, a.k.a. Adjusted R-squared ($R^2$-$adj$), around 0.912 which means that 91,2\% of variance present in dataset can be explained through this estimated model. In other words, this high $R^2$-$adj$ indicates that the model is well fitted to dataset used in building process which, in turn, allows it to perform more accurate predictions about the dependent variable. 


In addition, we assessed statistical significance of the regression model and coefficients of each term by applying hypothesis test F. At a level of significance ($\alpha$) of 0.05, test F revealed that regression as well as its coefficients can be considered as being statistically significant since resulted \textit{p-values} were smaller than 0.00. These results indicate that the probability of each coefficient has been estimated just for this sample is practically 0\%. In other words, this model has almost 100\% chance to be able to predict interference level for any sample besides that one used to build model. 

Moreover, an analysis accomplished on residuals showed that estimated model did not violate any of the MRA basic assumptions. Thus, residuals presented (i) linearity, (ii) homocedasticity and (iii) normal distribution.

\section{Experimental Tests and Results}

In this section, we describe experimental tests accomplished to assess the precision achieved by our model when predicting interference among co-located applications. In subsection \ref{sec:workload}, we describe the workload used for conducting such experimental tests. In subsection \ref{sec:predicting}, we present predictions made by using our model, while in subsection \ref{sec:precision} we evaluate how precisely these predictions match to interference levels achieved in real experiments. 

\subsection{Evaluation Workload}
\label{sec:workload}

In order to evaluate the precision of our prediction model, we accomplished interference experiments by using an evaluation workload comprised of a real petroleum reservoir simulator, called MUFITS, and applications from High Performance Computing Challenge Benchmark (HPCC). 

MUFITS is employed by petroleum engineers to study the behavior of petroleum reservoir across the time. From simulation results, they can make inferences about future conditions of the reservoir in order to maximize oil and gas production in a new or developed field. Basically, the simulator employs partial differential equations to describe the multiphase fluid flow (oil, water and gas) inside a porous reservoir rock \cite{peaceman2000fundamentals} \cite{alves2014}. Reservoir simulation is one of the most expensive computational problems faced by petroleum industry since a single simulation can take several days, even weeks, to finish. Computational complexity of this problem arises from the high spatial heterogeneity of multi-scale porous media \cite{lu2009iterative} \cite{yu2012gpu} \cite{habiballah2003parallel}. 

Besides MUFITS, we tested applications from HPCC, a widely adopted benchmark to evaluate performance of HPC systems. This benchmark provides seven kernels in total, but, only four of them, represent real HPC applications or operations commonly employed in scientific computing. A brief description of these four applications is as follows \cite{luszczek2006hpc} \cite{li2013evaluating} \cite{younge2011analysis} \cite{dongarra2013hpc}.    

\begin{itemize}
    \item HPL: solves a dense linear system of equations by applying the LU factorization method with partial row pivoting. This application, that is usually employed to measure sustained floating point rate of HPC systems, is the basis of evaluation for the Top 500 list. 
    \item DGEMM: performs a double precision real matrix-matrix multiplication by using a standard multiply method. Even not being a complex real application, this kernel represents one of the most common operation performed in scientific computing, the matrix-matrix multiplication.  
    \item PTRANS: performs a parallel matrix transpose. As their pairs of processors communicate with each other simultaneously, this application is a useful test to evaluate the total communications capacity of the network.
    \item FFT: computes a Discrete Fourier Transform (DFT) of very large one-dimensional complex data vector and is often used to measure floating point rate execution of HPC systems. 
\end{itemize}

We considered, for each of those applications, distinct instances in order to certify that our proposal is able to precisely predict interference, regardless of the size of instance treated by applications.  For MUFITS, we considered instances usually adopted in literature. The first one, labeled here as ``I1", concerns to a simulation of CO2 injection in the Johansen formation by using a real-scale geological model of the formation. The other one, labeled here as ``I2", is related to the 10th SPE (Society of Petroleum Engineers) Comparative Study. Both instances are available on MUFITS website\footnote{http://www.mufits.imec.msu.ru/}. 

For HPCC, we considered instances whose details are described in Table \ref{tab:instances}. We created these instances by adjusting parameters ``\#N", ``N" and ``NB" which correspond, respectively, to the number of problems, the size of the problem treated by application and the size of the block. Remark that each input parameter has a specific meaning for each application. For example, in case of DGEMM, input parameter ``N" is used to set the dimension of matrices to be multiplied, while this same parameter, in case of FFT, determines the size of vector of real numbers to be transformed to the frequency domain. More detailed information about these input parameters can be found in HPCC website\footnote{http://icl.cs.utk.edu/hpcc/}.      

\begin{table}[h]
\centering
\label{tab:instances}
\begin{tabular}{|c|c|c|c|c|}
\hline
\multirow{2}{*}{\textbf{Application}} & \multirow{2}{*}{\textbf{Instance}} & \multicolumn{3}{c|}{\textbf{HPCC Parameters}} \\ \cline{3-5} 
                                      &                                    & \textbf{\#N}   & \textbf{N}   & \textbf{NB}   \\ \hline
\multirow{2}{*}{HPL}                  & I1                                 & 1              & 18000        & 80            \\ \cline{2-5} 
                                      & I2                                 & 1              & 15000        & 80            \\ \hline
\multirow{2}{*}{DGEMM}                & I1                                 & 1              & 3000         & 80            \\ \cline{2-5} 
                                      & I2                                 & 1              & 18000        & 80            \\ \hline
\multirow{2}{*}{PTRANS}               & I1                                 & 1              & 24000        & 80            \\ \cline{2-5} 
                                      & I2                                 & 5              & 500          & 80            \\ \hline
\multirow{2}{*}{FFT}                  & I1                                 & 1              & 65000        & 10            \\ \cline{2-5} 
                                      & I2                                 & 1              & 40000        & 10            \\ \hline
\end{tabular}
\caption{HPCC applications instances description}
\end{table}


\subsection{Predicting Interference with Delfos}
\label{sec:predicting}

In order to assess model precision before distinct scenarios, we considered two co-locations schemes that resulted in 90 co-locations in total. In the first one, namely A, applications were co-located in a two-by-two fashion, where each application was executed by using instances I1 and I2. In this scenario, each one of those applications used 6 virtual machines except for FFT that used 4 virtual machines since this application restricts the number of process to a power of two. In the second co-location scheme, namely B,  those applications were co-located in a three-by-three fashion, where each application used 4 virtual machines. Unlike A, in scenario B each application solved just instance I1. Moreover, each virtual machine used in both schemes has the same configuration as the one described in subsection \ref{sec:synthetic} and, as performed in synthetic experiments, half of virtual machines allocated to each application was pinned to each NUMA node. 

For predicting interference suffered by applications when co-located on schemes A and B, we executed individually each of one of those applications in order to obtain their access rates to each shared resource. As described in Table \ref{tab:workload}, FFT achieved the highest access rate to virtual network, while PTRANS, when solving I1, imposed the highest pressure to SLLC and DRAM. Moreover, as DGEMM is a embarrassingly parallel application, it performed a very low access to virtual network. Remark, at last, that some applications decreased their access rates to shared resources when executing with four processors instead of six. This is expected since a same application, when executed with a lower number of processes, perform, usually, a lower amount of accumulated access to shared resources. 

\begin{table*}[]
\centering
\label{tab:workload}
\begin{tabular}{|c|c|c|c|c|c|c|}
\hline
\multirow{2}{*}{\textbf{Label}} & \multirow{2}{*}{\textbf{Application}} & \multirow{2}{*}{\textbf{Instance}} & \multirow{2}{*}{\textbf{\begin{tabular}[c]{@{}c@{}} Processes\end{tabular}}} & \multicolumn{3}{c|}{\textbf{Score}}          \\ \cline{5-7} 
                                &                                       &                                    &                                                                                            & \textbf{SLLC} & \textbf{DRAM} & \textbf{NET} \\ \hline
MUFITS.I1.P6                    & MUFITS                                & I1                                 & 6                                                                                          & 0.05          & 0.13          & 0.00         \\ \hline
MUFITS.I2.P6                    & MUFITS                                & I2                                 & 6                                                                                          & 0.03          & 0.00          & 0.01         \\ \hline
MUFITS.I1.P4                    & MUFITS                                & I1                                 & 4                                                                                          & 0.05          & 0.08          & 0.00         \\ \hline
HPL.I1.P6                       & HPL                                   & I1                                 & 6                                                                                          & 0.03          & 0.06          & 0.02         \\ \hline
HPL.I2.P6                       & HPL                                   & I2                                 & 6                                                                                          & 0.03          & 0.06          & 0.02         \\ \hline
HPL.I1.P4                       & HPL                                   & I1                                 & 4                                                                                          & 0.02          & 0.04          & 0.01         \\ \hline
DGEMM.I1.P6                     & DGEMM                                 & I1                                 & 6                                                                                          & 0.02          & 0.02          & 0.00         \\ \hline
DGEMM.I2.P6                     & DGEMM                                 & I2                                 & 6                                                                                          & 0.01          & 0.02          & 0.00         \\ \hline
DGEMM.I1.P4                     & DGEMM                                 & I1                                 & 4                                                                                          & 0.01          & 0.02          & 0.00         \\ \hline
PTRANS.I1.P6                    & PTRANS                                & I1                                 & 6                                                                                          & 0.18          & 0.21          & 0.32         \\ \hline
PTRANS.I2.P6                    & PTRANS                                & I2                                 & 6                                                                                          & 0.02          & 0.04          & 0.02         \\ \hline
PTRANS.I1.P4                    & PTRANS                                & I1                                 & 4                                                                                          & 0.14          & 0.09          & 0.19         \\ \hline
FFT.I1.P4                       & FFT                                   & I1                                 & 4                                                                                          & 0.07          & 0.17          & 0.49         \\ \hline
FFT.I2.P4                       & FFT                                   & I2                                 & 4                                                                                          & 0.07          & 0.16          & 0.52         \\ \hline
\end{tabular}
\caption{Individual profiles of applications used in experiments}
\end{table*}



Considering individual profile of each application, we applied our quantitative model to predict what would be the interference suffered by those applications when co-located in accordance with schemes A and B. Prediction results point out that the minimum and maximum predicted interference levels would be equal to 1.53\% and 43.10\%, respectively. As expected, the lowest and highest interference levels were predicted for co-locations that involved, respectively, applications with low and high access rates to SLLC, DRAM and virtual network. In other words, our model indicated that DGEMM and HPL would suffer a low cross-interference, while FFT and PTRANS would present the highest interference levels. 

However, specifically to PTRANS, our model point out that interference suffered by this application would vary significantly depending on instance being solved. Our model indicated that PTRANS.I1.P6 would experience an interference level around 40\% when co-located with itself in a two-by-two fashion, while PTRANS.I2.P6, when co-located in same conditions, would present an interference level of approximately 12\%. 

Moreover, our model predicted that PTRANS and DGEMM, although belonging to the same Dwarf class, namely Dense Linear Algebra, would present distinct interference levels when co-located with themselves. So, prediction results indicated that interference level resulted from co-location ``PTRANS.I1.P6xPTRANS.I1.P6" would be approximately equal to 40\%, while "DGEMM.I1.P6xDGEMM.I1.P6" would suffer a interference level close to 3\%. 

Furthermore, our quantitative model predicted that FFT.I1.P4 and MUFITS.I1.P4, although have presented similar SLLC access rates, would suffer distinct interference levels when co-located with themselves in a three-by-three fashion. Specifically, our solution predicted that FFT.I1.P4 would present a mutual interference around 40\%, while MUFITS.I1.P4 would suffer a cross-interference approximately equal to 12\%.

\subsection{Evaluating Precision of Interference Prediction}
\label{sec:precision}

In order to evaluate the precision of our quantitative model, we executed all of the co-locations defined in schemes A and B and, for each co-location, we calculated the \textit{prediction error} achieved by our solution. The prediction error is defined as the absolute value of the difference between interference level predicted by our model and the real interference level suffered by applications.

\begin{figure}[h] 
    \centering
    \includegraphics[scale=0.47]{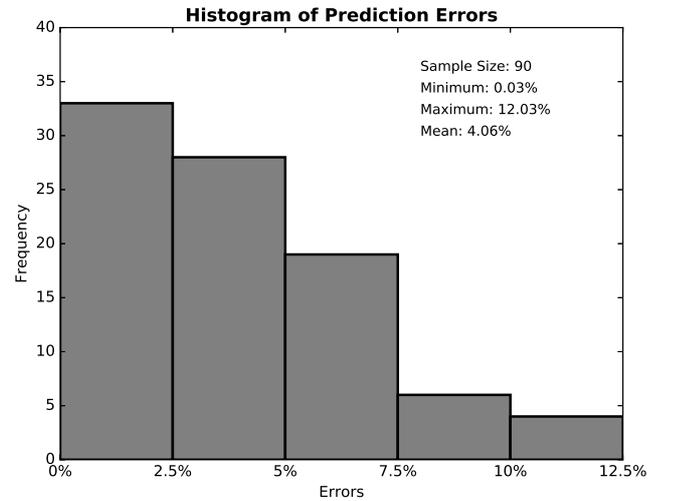}
    \caption{Histogram of prediction errors}
    \label{fig:predict_results}
\end{figure}

As can be seen in Figure \ref{fig:predict_results}, our model presented an average and maximum prediction errors equal to 4.06\% and 12.03\%, respectively. Moreover, in approximately 96\% of all tested cases, our quantitative model presented a prediction error less than 10\%. Such results revealed that our model, although have been built from a two-by-two fashion experiment, precisely predicted cross-interference for cases which three applications were executed simultaneously, as well. 

Some interesting results are highlighted in Table \ref{tab:results_prediction}. At first, as predicted by our model, experimental results showed that PTRANS really presented distinct interference levels when treating I1 and I2. Our quantitative model was able to predict interference of PTRANS regardless of the size of instance because it takes into account the amount of access that applications perform to shared resources. Indeed, as can be seen in Table \ref{tab:workload}, the amount of access that PTRANS performs to all shared resources when solving I1 is significantly higher than the one achieved when treating I2. 

\begin{table*}[]
\centering

\label{tab:results_prediction}
\begin{tabular}{|c|c|c|c|}
\hline
\multirow{2}{*}{\textbf{Co-location}}  & \multicolumn{2}{c|}{\textbf{Interference Level}} & \multirow{2}{*}{\textbf{\begin{tabular}[c]{@{}c@{}}Prediction \\ Error\end{tabular}}} \\ \cline{2-3}
                                       & \textbf{Real}        & \textbf{Predicted}        &                                                                                       \\ \hline
PTRANS.I1.P6xPTRANS.I1.P6              & 44.50\%              & 39.97\%                   & 4.53\%                                                                                \\ \hline
PTRANS.I2.P6xPTRANS.I2.P6              & 5.31\%               & 12.11\%                   & 6.80\%                                                                                \\ \hline
DGEMM.I1.P6xDGEMM.I1.P6                & 7.79\%               & 2.50\%                    & 5.29\%                                                                                \\ \hline
FFT.I1.P4xFFT.I1.P4xFFT.I1.P4          & 49.31\%              & 40.45\%                   & 8.87\%                                                                                \\ \hline
MUFITS.I1.P4xMUFITS.I1.P4xMUFITS.I1.P4 & 22.85\%              & 11.65\%                   & 11.20\%                                                                               \\ \hline
\end{tabular}
\caption{Subset of interference experiments}
\end{table*}


Besides that, those experimental results confirmed that PTRANS and DGEMM, even belonging to the same Dwarf class, presented distinct interference levels. This result allows to state that a qualitative approach based on Dwarfs classes is not enough to precisely determine interference. On the other hand, our model was able to predict that PTRANS and DGEMM would present, respectively, a high and a low interference levels.  

At last, results also showed that SLLC access contention is not the only cause for the cross-interference problem. FFT.I1.P4 and MUFITS.I1.P4, although have similar SLLC access burden, suffered distinct interference levels. As depicted in Table \ref{tab:results_prediction}, our model predicted satisfactorily interference suffered by these applications because it evaluates not only SLLC, but DRAM and virtual network access as well. Indeed, although both applications present similar SLLC access rates, FFT.I1.P4, which suffered a higher interference, performs a higher access to virtual network than MUFITS.I1.P4. 

Therefore, all these findings allow to assert that a solution based on Dwarfs classes or that evaluates just SLLC access contention is not suitable to determine interference suffered co-located applications. Our proposed model satisfactorily predicted interference for these specific cases because it takes into account the amount of simultaneous access to SLLC, DRAM and virtual network, besides considering access burden similarities among co-located applications.

\section{Conclusions and Future Work}

In this paper, we presented a quantitative model that takes into account the amount of simultaneous access to SLLC, DRAM and virtual network, and the similarity of application's access burden to predict the level of interference suffered by applications when sharing a same physical machine. Experimental results, considering a real petroleum reservoir simulator and applications from HPCC benchmark, showed that our solution was able to predict interference with an average and maximum errors around 4\% and 12\%, respectively. Besides that, in 96\% of all tested cases, our solution reached a  prediction error less than 10\%. 

More specifically, our experimental tests showed that our solution could correctly predict interference of co-located applications even for cases which interference suffered by the same application varied before distinct instances. In addition, our model was able to precisely predict interference regardless the number of applications being co-executed in same host, though it was built by using a dataset generated from a two-by-two fashion experiment. Furthermore, our model could predict interference for co-locations that, even presenting similar SLLC access burden, achieved distinct interference levels. Our model accurately predicted interference in this case because it considers, besides SLLC, other shared resources such as DRAM and virtual network.

In future works, we expect to evaluate the influence that concurrent access performed to other shared resources, like disk, impose in cross-application interference. Moreover, we are also interested in investigating whether the total amount of allocated memory has a direct impact in interference suffered by co-located applications, besides assessing other metrics of virtual network such as the number of packets transmitted, for example. Furthermore, it would be interesting to evaluate whether interference varies before distinct hardware configuration in order to incorporate this issue to our prediction model.

\bibliographystyle{elsarticle-harv} 
\bibliography{interference}

\end{document}